\documentclass[3p,times,twocolumn]{elsarticle}

\usepackage{ecrc}

\usepackage{graphicx}
\usepackage{caption}
\usepackage{subcaption}
\usepackage{placeins}

\volume{00}

\firstpage{1}

\journalname{Nuclear Physics B Proceedings Supplement}

\runauth{C. Bigongiari for the CTA Consortium}

\jid{nuphbp}

\jnltitlelogo{Nuclear Physics B Proceedings Supplement}

\usepackage{amssymb}

\usepackage[figuresright]{rotating}

\begin{document}

\begin{frontmatter}



\dochead{}

\title{The Cherenkov Telescope Array}


\author[OATo]{Ciro Bigongiari for the CTA Consortium}

\address[OATo]{Osservatorio Astrofisico di Torino, Strada Osservatorio 20, 10025, Pino Torinese (To), Italy}

\begin{abstract}
The Cherenkov Telescope Array (CTA) 
is planned to be the next generation ground based
observatory for very high energy (VHE) gamma-ray astronomy. 
Gamma-rays provide a powerful insight into the non-thermal universe 
and hopefully a unique probe for new physics. 
Imaging Cherenkov telescopes have already discovered more than 170 
VHE gamma-ray emitters providing plentiful of valuable data 
and clearly demonstrating the power of this technique. 
In spite of the impressive results there are indications that the known sources
represent only the tip of the iceberg.
A major step in sensitivity is needed 
to increase the number of detected sources, observe short time-scale variability 
and improve morphological studies of extended sources. 
An extended energy coverage is advisable to observe far-away extragalactic objects and 
improve spectral analysis. 
CTA aims to increase the sensitivity by an order of magnitude 
compared to current facilities, to extend the accessible gamma-ray energies from a few
tens of GeV to a hundred of TeV, and to improve on other parameters like angular and energy
resolution.
CTA will provide moreover a full sky-coverage 
by featuring an array of imaging atmospheric Cherenkov telescopes 
in both hemispheres.
%
%
%

This paper presents an overview 
of the technical design and summarize the current
status of the project. 
CTA prospects 
for some key science topics like the origin of
relativistic cosmic particles, 
the acceleration mechanisms 
in extreme environments such as neutron
stars and black holes 
and searches for Dark Matter are discussed. 

\end{abstract}


\end{frontmatter}


\section{Introduction}
\label{INTRO}

The Cherenkov Telescope Array (CTA) \cite{CTA1,CTA2} is an ongoing project to 
build an observatory for gamma ray astronomy operating in the Very High Energy (VHE) 
range, between few tens of Giga-electron-volts (GeV) to few hundreds of 
Tera-electron-volts (TeV).  
The consortium carrying on this project is presently composed of  
1200 scientists and engineers working in 200 institutes from 32 countries.  
The CTA consortium adds up all the main research groups in this field 
resulting in an unprecedented convergence of human resources and know-how.

The CTA observatory will dramatically outperform present generation arrays of Imaging Air 
Cherenkov Telescopes (IACT), 
like H.E.S.S. \cite{HESS}, MAGIC \cite{MAGIC} and VERITAS \cite{VERITAS} 
in all aspects of performance.
A ten fold improvement in flux sensitivity, a much wider energy coverage and
greatly improved energy and angular resolutions are expected.  
CTA telescopes will have also a larger field of view to ensure much better surveying capabilities.    
Such a performance guarantees very significant scientific return, in the form
of precision very-high-energy astrophysics, and
considerable potential for major discoveries in astrophysics and fundamental physics.

The CTA observatory will detect more than one thousand
VHE sources, ten times more than currently known, 
allowing for population studies of source classes, like 
Pulsar Wind Nebulae or Active Galactic Nuclei. 

Thanks to its higher sensitivity and better angular resolution 
CTA will perform very detailed morphological studies of extended 
sources, while its wider energy coverage and improved energy resolution 
will allow spectral studies of unprecedented accuracy. 

The CTA performance will also greatly improve the
chances for the detection of new phenomena and for 
discoveries in fundamental physics issues such as 
Dark Matter,  Lorentz invariance violation at Planck scale, 
extragalactic background light (EBL) or axion physics. 

CTA  will be the first astro-particle
instrument to be operated as an open
observatory, providing a wealth of data to a wide astrophysics community. 
CTA aims moreover to become a cornerstone in a networked multi-wavelength, 
multi-messenger exploration of high-energy astrophysical phenomena.
 
Details on the science cases and on technical implementations of
CTA can be found in \cite{CTAscience,CTAproject}.

\section{Gamma-ray astronomy}
\label{GRA}

Satellite-borne detectors like Agile \cite{AGILE}, Fermi \cite{FERMI} and Integral \cite{INTEGRAL}
are sensitive to high energy gamma-rays ($\sim 30$~MeV - $\sim 30$~GeV). 
At higher energies the gamma-ray fluxes from cosmic astrophysical sources are so low 
that such detectors cannot collect a statistically significant 
number of events in a reasonable amount of time.    

\begin{figure*}[htb]
  \centering
  \includegraphics[width=\textwidth]{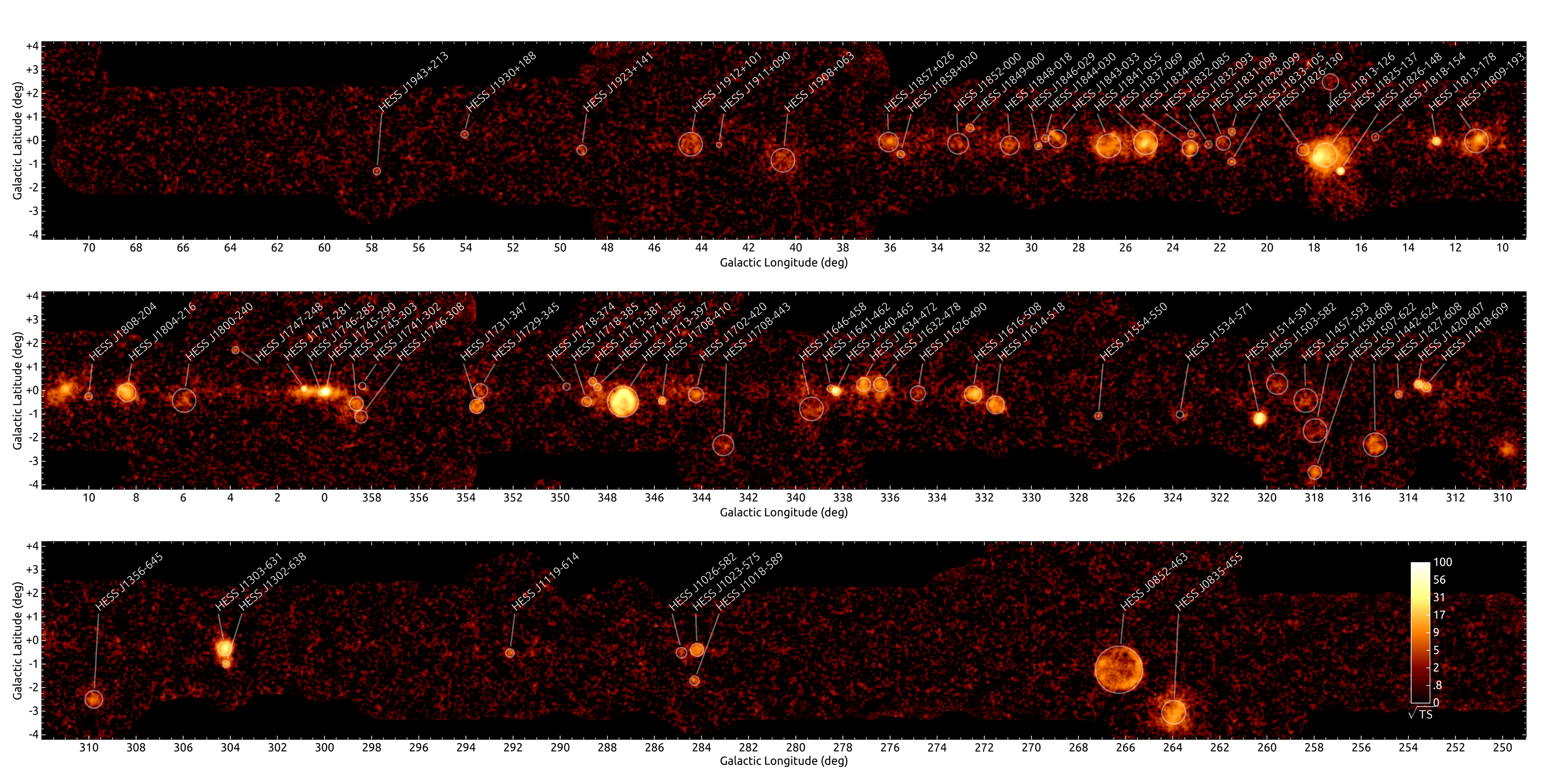}
  \caption{TeV source map for the H.E.S.S. Galactic Plane Survey \cite{HESS_GalacticPlaneSurvey}. 
    Identifiers for sources that have been described in publications or announced at conferences are included. 
    }
  \label{fig:HESS_GP}
\end{figure*}

Very high energy gamma-rays impinging on the Earth, interact with
atmospheric nuclei usually producing an electron-positron pair 
which in turn generates a cascade of particles, 
mostly electrons and positrons, usually called electromagnetic shower. 
The showers develop down in the atmosphere extending for several kilometers
while spreading few hundreds of meters apart.
All the shower particles for primaries up to about $\sim 100$~TeV 
stop high up in the atmosphere, and so cannot be directly detected 
by ground based detectors like HAWC \cite{HAWC} or the upcoming LHAASO \cite{LHAASO}. 
However, a sizable fraction of the shower particles, 
travels at ultra-relativistic speed and emits Cherenkov light. 
The near UV and optical component of such radiation can propagate 
nearly unattenuated to the ground, with
minor losses due to Rayleigh and Mie scattering and Ozone absorption. 
IACTs reflect the collected Cherenkov light onto the focal plane 
where a multi-pixel camera records the shower image.
The shape, size and orientation of the image provide valuable information 
about the primary energy and its direction of propagation.  
Moreover such information can be used to reduce the overwhelming background 
due hadronic primaries which produce wider and more irregular images.  
This technique, pioneered by the Whipple collaboration, 
which first detected the Crab Nebula at VHE, in 1989 \cite{WhippleCrab}, 
has proven very successful leading to the detection of more than 170 sources in the last 25 years \cite{TEVCAT}
and to many exciting results.  
H.E.S.S. realized the first deep survey in VHE gamma rays of the Milky Way discovering many new sources, 
in particular pulsar wind nebulae which are the most common galactic objects to emit gamma rays, see figure \ref{fig:HESS_GP} 
\cite{HESS_GalacticPlaneSurvey}.

Deep observations of selected objects enabled detailed morphological studies of extended sources like
the young supernova remnant RX J1713.7-3946 \cite{HESS_RX_J1713} which provide valuable information about 
emission regions and acceleration mechanisms. 
Many extragalactic sources have been detected too,
mostly blazars which have been observed up to redshift $z=0.944$ \cite{S3_0218}.
Some other classes of extragalactic objects have been confirmed as VHE gamma-rays emitters like radio galaxies, 
Flat Spectrum Radio Quasars and star-bust galaxies. 
For recent review of the status of the gamma-ray astronomy see, e.g., \cite{GammaRayReview}.

\section{Improvements for the next generation of imaging air Cherenkov telescopes}
\label{NextIACT}

Regardless of the very exciting achievements of current-generation Cherenkov telescopes, there are
limitations that next generation instruments should overcome: current IACTs are sensitive
in an energy range of 80~GeV - 50~TeV, have a typical full Field of View (FoV) of $3^{\circ}$ - $5^{\circ}$, 
their angular resolution is currently around some arcmin above 1~TeV and their energy resolution approaches $10\%$ 
well above the threshold.  
From a physics point of view, there are strong arguments to decrease the energy threshold
to few tens of GeV, acquire sensitivity beyond 50~TeV and increase sensitivity in the core range
(100~GeV − 50~TeV). 
Extending the observations well below 80 GeV will be crucial for many studies about galactic sources, 
providing for example the final answer about acceleration mechanisms in pulsars, 
whose energy spectrum is expected  to have a cut-off just in this region. 
Lowering the energy threshold will be extremely important also for the observation of distant extragalactic sources 
because their gamma-ray flux at higher energies is highly attenuated 
by the interaction with the Extragalactic Background Light (EBL). 
An increased sensitivity in the core energy region along with a larger FoV will allow for a full VHE sky survey, 
with more than one thousand new gamma-ray sources expected to be detected.
Moreover a sub-minute time resolution will be achievable with a ten-times better sensitivity 
allowing more detailed observation of variable sources like gamma-ray outburst by Active Galactic Nuclei. 
Extending the observation above 50~TeV will allow instead to understand the acceleration mechanism 
in galactic objects like SNRs, 
discriminating between hadronic and leptonic models. 
This energy range will be also important for identifying the so-called PeVatrons, 
i.e. accelerators of cosmic rays up to PeV energies.
A wider FoV will improve the surveying capabilities, ease the study of extended sources and increase the chance to observe 
two sources at the same time. 
A better angular resolution will increase the rejection of hadronic background improving the sensitivity, will provide 
finer emission maps of extended sources and will reduce source confusion improving collaboration with instruments 
observing at other wavelengths.   
All the spectral studies to constrain acceleration mechanisms or to search for Dark Matter signals will benefit from an 
improved energy resolution 
The wish-list for the next generation gamma-ray detectors can therefore be summarized as: 

\begin{itemize}
  \item Decrease the energy threshold to few tens of GeV. 
  \item Extend the sensitivity range up to 100~TeV.
  \item Increase the sensitivity in the core energy region by at least an order of magnitude. 
  \item Improve the angular resolution down to few arc-minutes. 
  \item Widen the telescope field of view. 
  \item Improve the energy resolution.  
\end{itemize}

\section{CTA conceptual design}
\label{CTAdesign}

The CTA project aims at implementing all the improvements discussed in the previous section 
and the experience gained with the current generation of IACTs 
clearly demonstrates that such goals are achievable with available technical solutions. 
A huge array of telescopes like those of H.E.S.S.-I or Veritas  
would meet these requirements but would be too expensive. 
The same performance can be achieved in a cost-effective way with a mixed array 
composed of different types of telescopes. 
Gamma-rays at the lower end of the desired energy range produce a very low density of Cherenkov photons on ground. 
They can be detected only by telescopes with a reflecting surface large enough to collect a number of Cherenkov 
photons suitable for the proper reconstruction of their images. 
At the upper end of the energy range instead, gamma-rays produce Cherenkov pulses so intense that they can be easily 
detected by small size telescope too. The gamma-ray flux at such energies is so low, even for strong sources,  
that a very large instrumented area 
is needed to collect a statistically significant amount of events in a reasonable observation time.   
In the core energy range gamma-rays can be efficiently detected by telescopes 
similar to H.E.S.S.-I or Veritas ones, but a large number of such telescopes is needed to increase the collection area 
and the fraction of events reconstructed by many telescopes, which provide 
better energy and angular resolution.  
An improved sensitivity over the full energy range as well as improved energy and angular resolution 
can be achieved therefore by combining many telescopes of three different sizes
distributed over a large area: few large size telescopes (LST),
several medium size telescopes (MST),  and many small size telescopes (SST).

On the basis of the previous considerations the CTA design has been developed on few basic ideas: 

\begin{itemize}

\item{Increase the array size up to $\sim100$ telescopes;}
\item{Distribute telescopes  over a large area ($1 − 10 \; \mathrm km^{2} $);}
\item{Make use of telescopes of at least three different sizes;}
\item{Take advantage of well-proven technology of current IACTs;}
\item{Increase automation and remote operation;}

\end{itemize}

During the CTA design phase many parameters concerning the mix of telescopes, 
the telescope optical designs as well as cameras and readout electronics 
have been optimized under constraints given by cost limits and reliability/durability requirements. 
According to the current design CTA will comprise three different kind of telescopes with the following 
main characteristics: 

\subsection{Large Size Telescopes}
The current design for LSTs consists of a parabolic dish of 23~m
diameter with f/D = 1.2 constructed using a carbon fiber structure according to the approach 
successfully followed by MAGIC. These telescopes will be equipped with PMT-based cameras
with $4.4^{\circ}$ full field of view. See top left panel of figure \ref{fig:CTAtelescopetypes}.    

\subsection{Medium Size Telescopes}

The MST baseline design is based on the well-proven experience
of H.E.S.S. and VERITAS collaborations: 
a 12~m diameter Davies-Cotton \cite{DC} reflector with f/D = 1.3
and a PMT-based camera covering at least $7^{\circ}$ FoV. 
An alternative design based on the Schwarzschild-Couder \cite{SC} dual-mirror
optical layout and a SiPM-based camera is being developed. 
See top and bottom right panels of figure \ref{fig:CTAtelescopetypes}.

\subsection{Small Size Telescopes}
\begin{figure*}[htb]
  \centering
  \begin{subfigure}[b]{0.35\textwidth}
    \includegraphics[width=\textwidth]{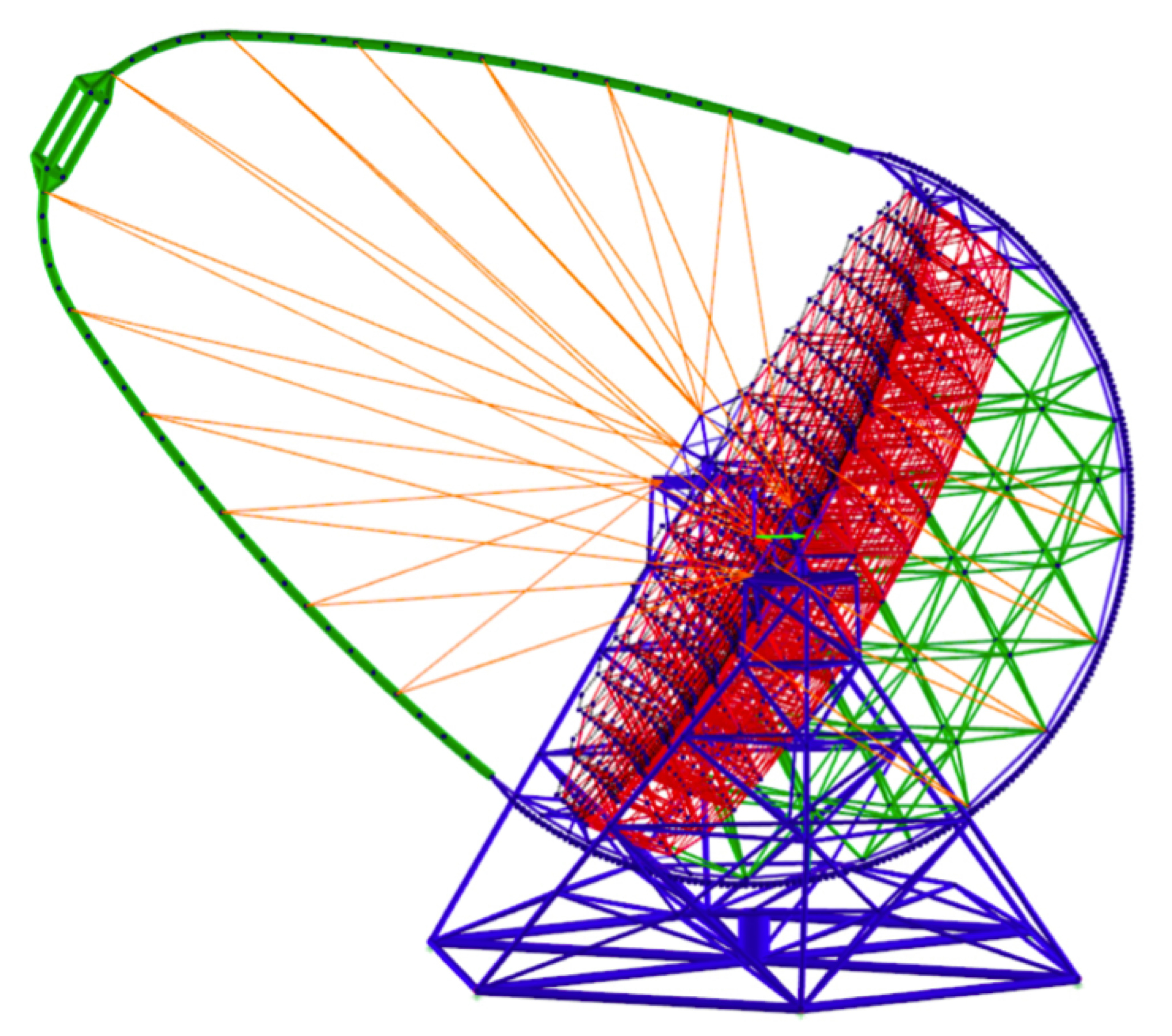}
  \end{subfigure}
  \hspace{3cm}
  \begin{subfigure}[b]{0.35\textwidth}
    \includegraphics[width=\textwidth]{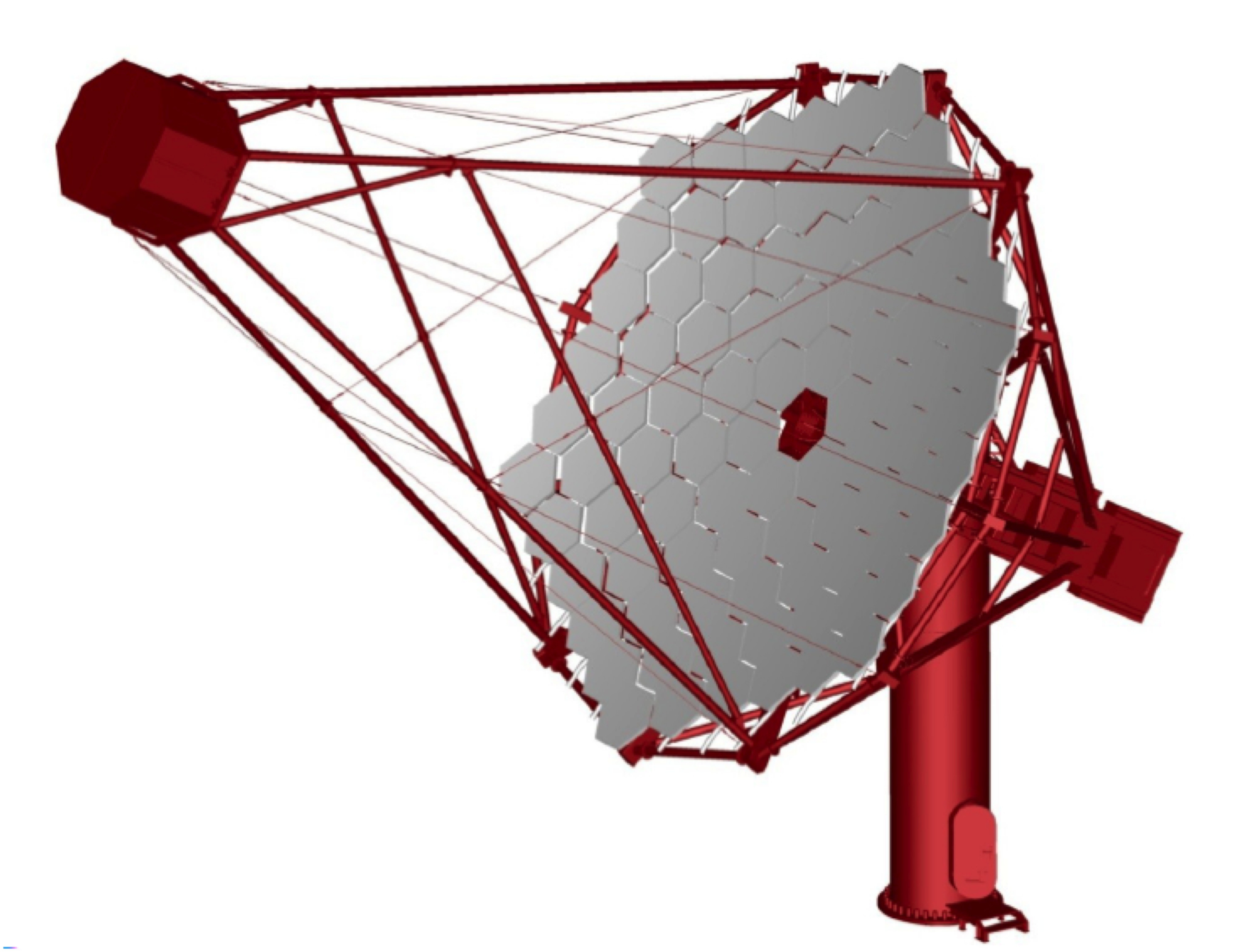}
  \end{subfigure}
  
  \begin{subfigure}[b]{0.35\textwidth}
    \includegraphics[width=\textwidth]{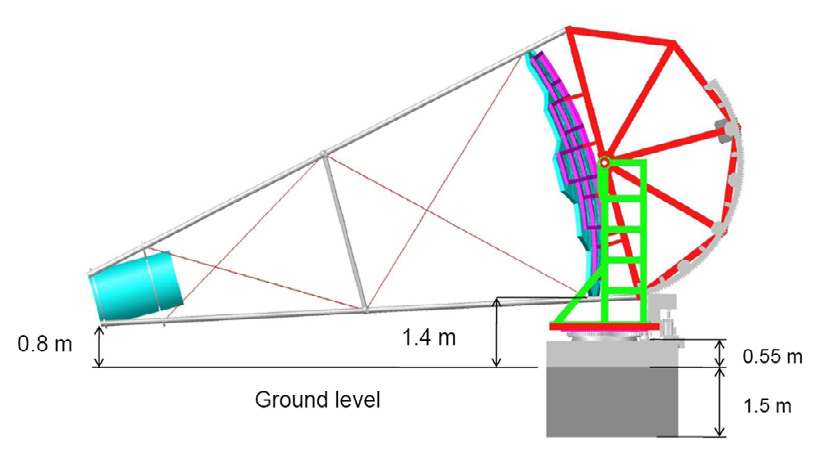}
  \end{subfigure}
  \hspace{3cm}
  \begin{subfigure}[b]{0.25\textwidth}
    \includegraphics[width=\textwidth]{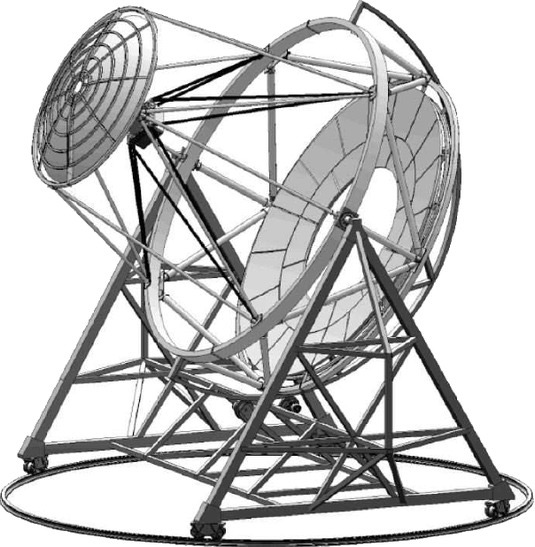}
  \end{subfigure}
  
  \caption{Top left: concept of a 23 m diameter LST with parabolic dish and f/D = 1.2. 
    Top right: concept of a 12 m diameter MST with a Davies-Cotton dish and f/D = 1.3. 
    Bottom left: concept of a 6 m diameter SST with a Davies-Cotton dish and f/D = 1.4. 
    Bottom right: concept of a dual mirror Schwarzschild-Couder MST telescope.}
  \label{fig:CTAtelescopetypes}
  
\end{figure*}

There are presently three different projects for the 
small size telescopes, one based on the Davies-Cotton optical 
design and two on the Schwarzschild-Couder. 
The primary mirror has a diameter of 5-6~m for all of them 
and the camera covers nearly $10^{\circ}$. 
See bottom left panel of figure \ref{fig:CTAtelescopetypes}.

\section{CTA sites}
\label{CTAsites}

In order to provide full sky coverage, 
the CTA observatory will consist of two arrays,  
one in the southern hemisphere covering the full energy range
and a second one in the northern hemisphere limited to few tens of TeV.  
The first one will provide detailed investigations of galactic sources, 
and in particular of the Galactic center, as well as
the observation of southern extragalactic objects while 
the second array will be mainly dedicated to northern extragalactic objects.
Many sites from several countries have been evaluated and two of them have been chosen for final negotiations in June 2015, the Observatorio del Roque de Los Muchachos on the La Palma island (Spain) for the northern array, 
and a site close to Paranal Observatory (Chile) for the southern array. 

\section{CTA arrays}
\label{CTAarrays}

The southern array will be composed of the three kinds of telescope described above  
to cover the full energy range, while the northern one will comprise only
large and medium size telescopes with a significant sensitivity up to 50~TeV.
The northern array will consist of 4 LSTs and 15 MSTs while the southern one 
will be composed of 4 LSTs, 25 Davies-Cotton MSTs and 70 SSTs. 
An extension for the southern observatory is envisaged based on a fourth,
medium-sized Schwarzschild Couder telescope type.

\begin{figure*}[ht]
  \centering
  \includegraphics[width=0.9\textwidth]{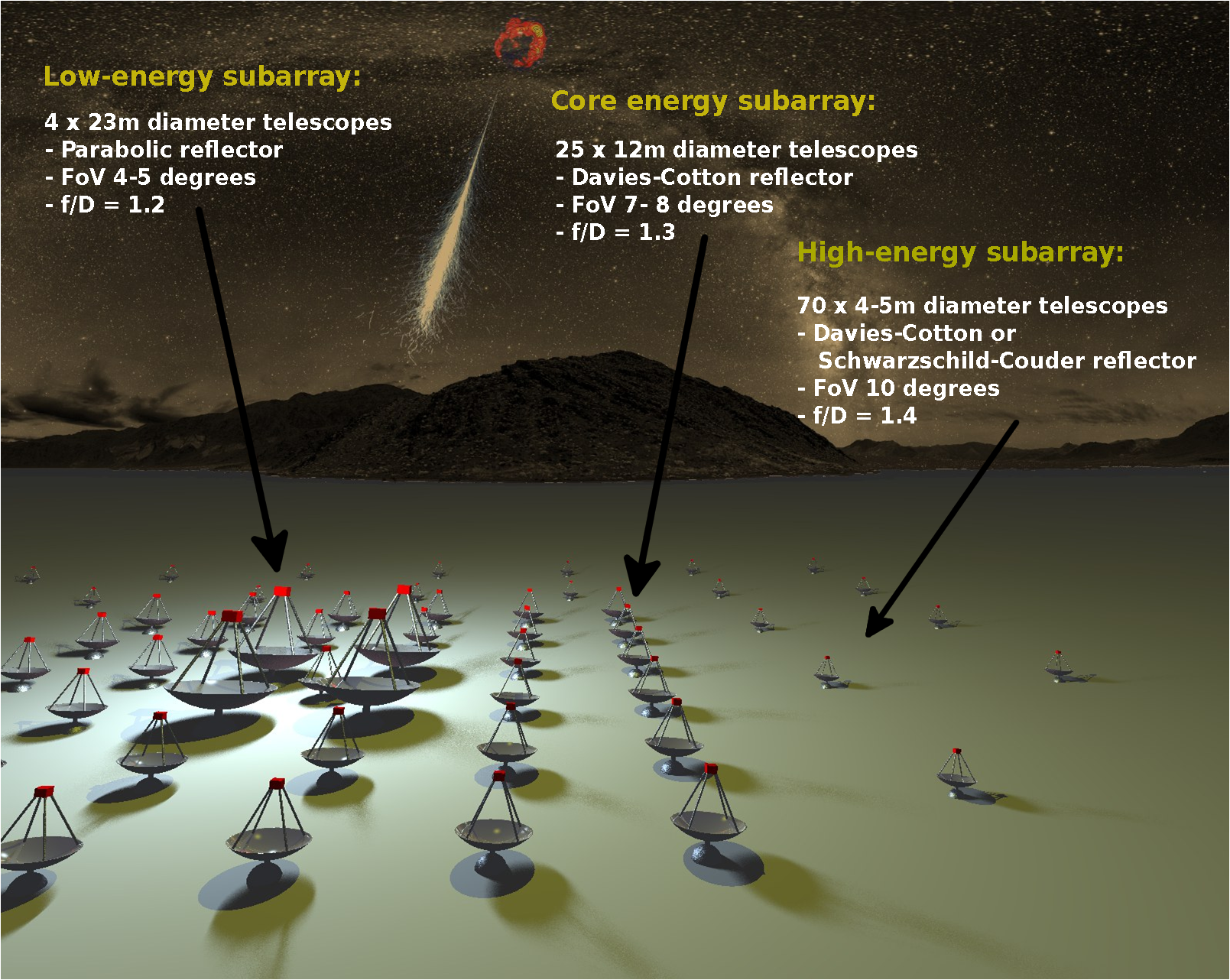}
  \caption{Schematic view (not to scale) of the baseline layout of the
CTA south array consisting of 3 types of telescopes. 
The details given in the sketch for each telescope type are discussed in reference \cite{CTAproject}.}

  \label{fig:CTAlayout}
\end{figure*}

The design of the CTA layout for the southern site is artistically depicted in Figure \ref{fig:CTAlayout}.

\section{CTA Expected Performance}
\label{CTAperformance}

\begin{figure}[t]
  \centering
  \includegraphics[width=0.45\textwidth]{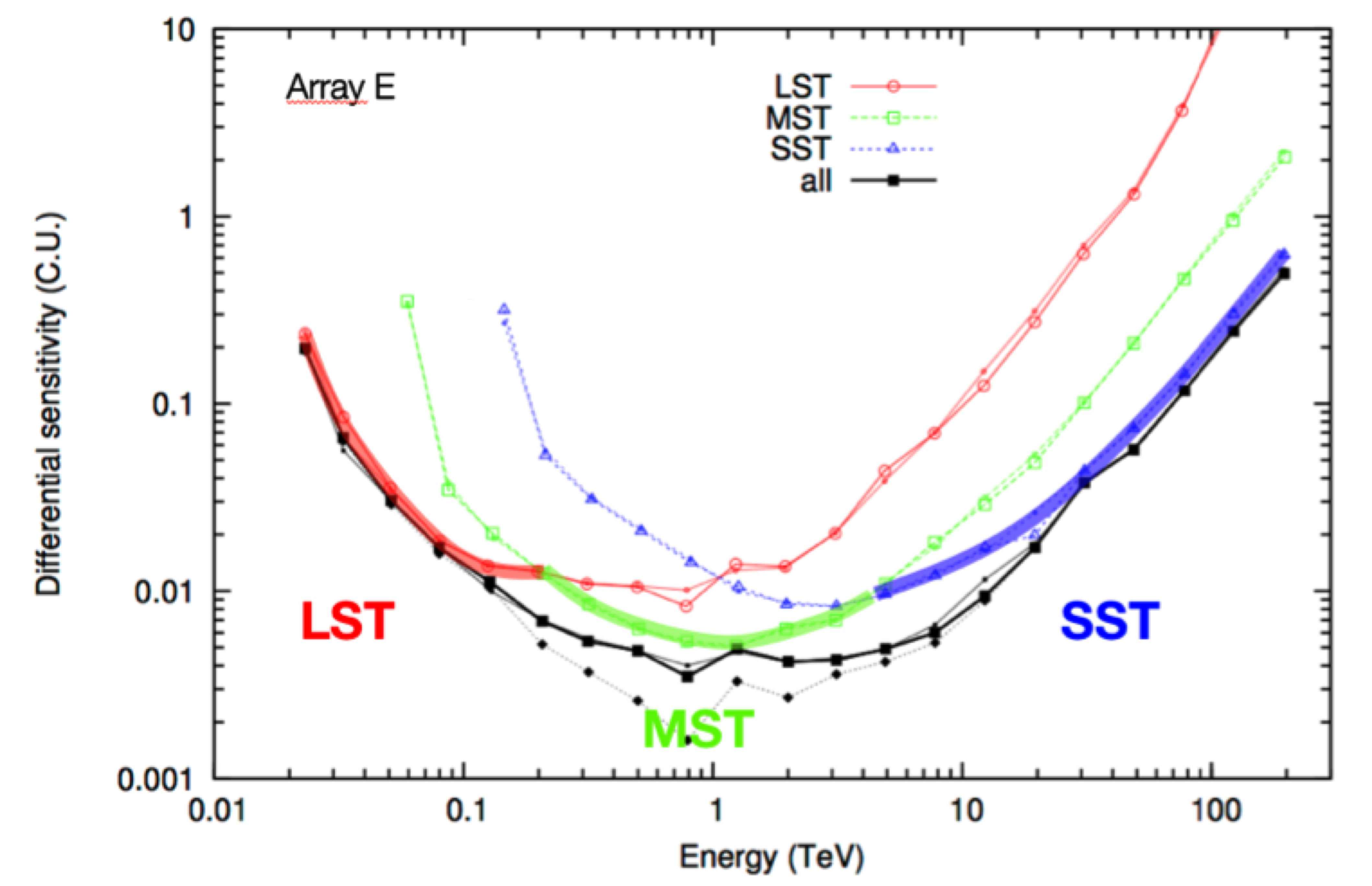}
  \caption{Point source sensitivity of for an example array layout
    (solid black line, filled squares) compared to that of its
    component sub-systems: 3 LSTs (red, open circles), 18 MSTs
    (green, open squares), and 56 SSTs (blue, open triangles).
    The importance of the background of cosmic ray electrons for the
    combined sensitivity is demonstrated by the dashed line with diamonds, where this background is ignored.
  }
  \label{fig:CtaDiffSens}
\end{figure}

The expected performance of CTA has been
evaluated with very detailed Monte Carlo simulations
for many different layouts and all considered sites \cite{CTAsimulations}. 
Atmospheric showers have been simulated using the well-known CORSIKA code \cite{CORSIKA}
while the atmospheric extinction and the telescope response have been simulated with the sim\_telarray package \cite{SIMTELARRAY}
which has been extensively verified by comparison to existing instruments. 
The required performance can be achieved at all CTA
candidate sites investigated and for a broad range of
implementation options, being rather insensitive to details of the telescope layout.
The contribution of the different types of telescopes
to the overall sensitivity for on-axis point sources is
showed in figure \ref{fig:CtaDiffSens}. 

The overall differential sensitivity for a 5$\sigma$ detection in 50~h of observation time with at least 10 signal events,  
reaches a minimum of few milli C.U.  
\footnote{1 C.U. (Crab Unit) = 
$1.5 \times 10^{3} \cdot (E/GeV)^{-2.58} \mathrm{photons\; cm^{-2} s^{-1} TeV^{-1}} $}
at a core energy around 1 TeV and stays well below 1 C.U. over the entire energy range. 
The cross-over in sensitivity between the LST and MST components occurs
at about 250~GeV and that between MSTs and SSTs at about 4 TeV. 
It's worthwhile to notice that at the 
transition points, the combined sensitivity is almost a factor of
two better than that of the individual components.

\begin{figure}[htb]
  \centering
  \includegraphics[width=0.50\textwidth]{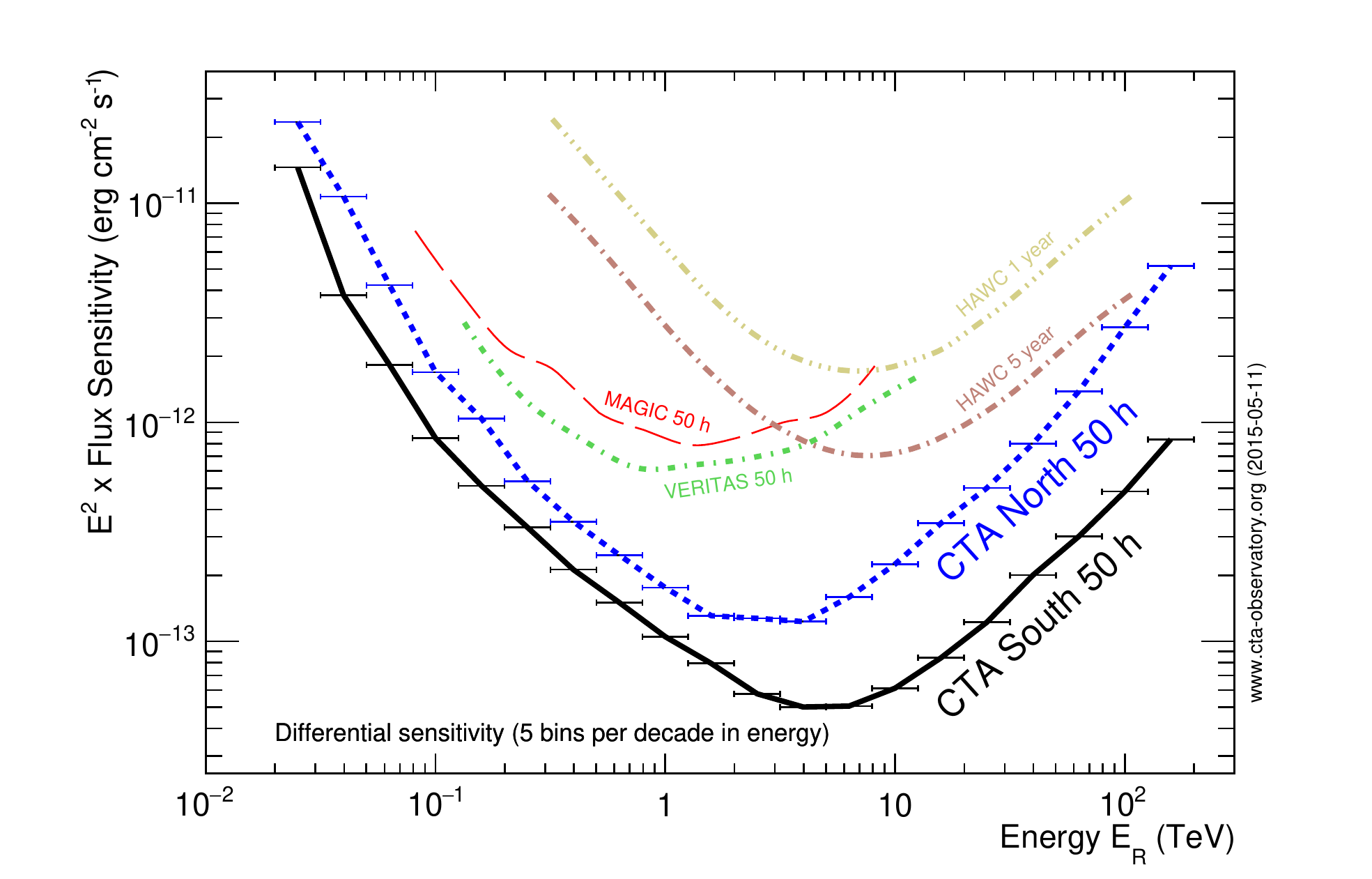}
  \caption{Expected differential sensitivity for CTA southern array (black line) 
  	and CTA northern array (blue line) \cite{CTAperf}. For comparison MAGIC-II \cite{MAGICperf} and VERITAS \cite{VERITASperf} sensitivities for the same observation time are shown (red line and green line respectively),  
    together with HAWC sensitivity \cite{HAWCperf} for an observation time of one year (yellow line) and five years (dark-red line).  
    }
  \label{fig:CtaSensComp}
\end{figure}

The CTA differential sensitivity for both southern and northern arrays for an observation time of 50 hours is compared in figure 
\ref{fig:CtaSensComp} 
with the sensitivities of MAGIC-II and Veritas  for the same observation time. 
HAWC sensitivities for an observation time of one year and five years 
are shown too for comparison. 
The sensitivity of the CTA northern array will be worse than that of the southern one 
over the entire energy range due to the lower number of telescopes. 
At energies above few TeVs the sensitivity of the CTA northern array will be further limited  
by the lack of SST telescopes. 
Nevertheless both arrays will outperform present IACT facilities and also last generation 
ground-based gamma-ray detectors like HAWC over the full energy range comprised by few tens of GeV and 100 TeV.    


\begin{figure}[hbt]
  \centering
   \includegraphics[width=0.5\textwidth]{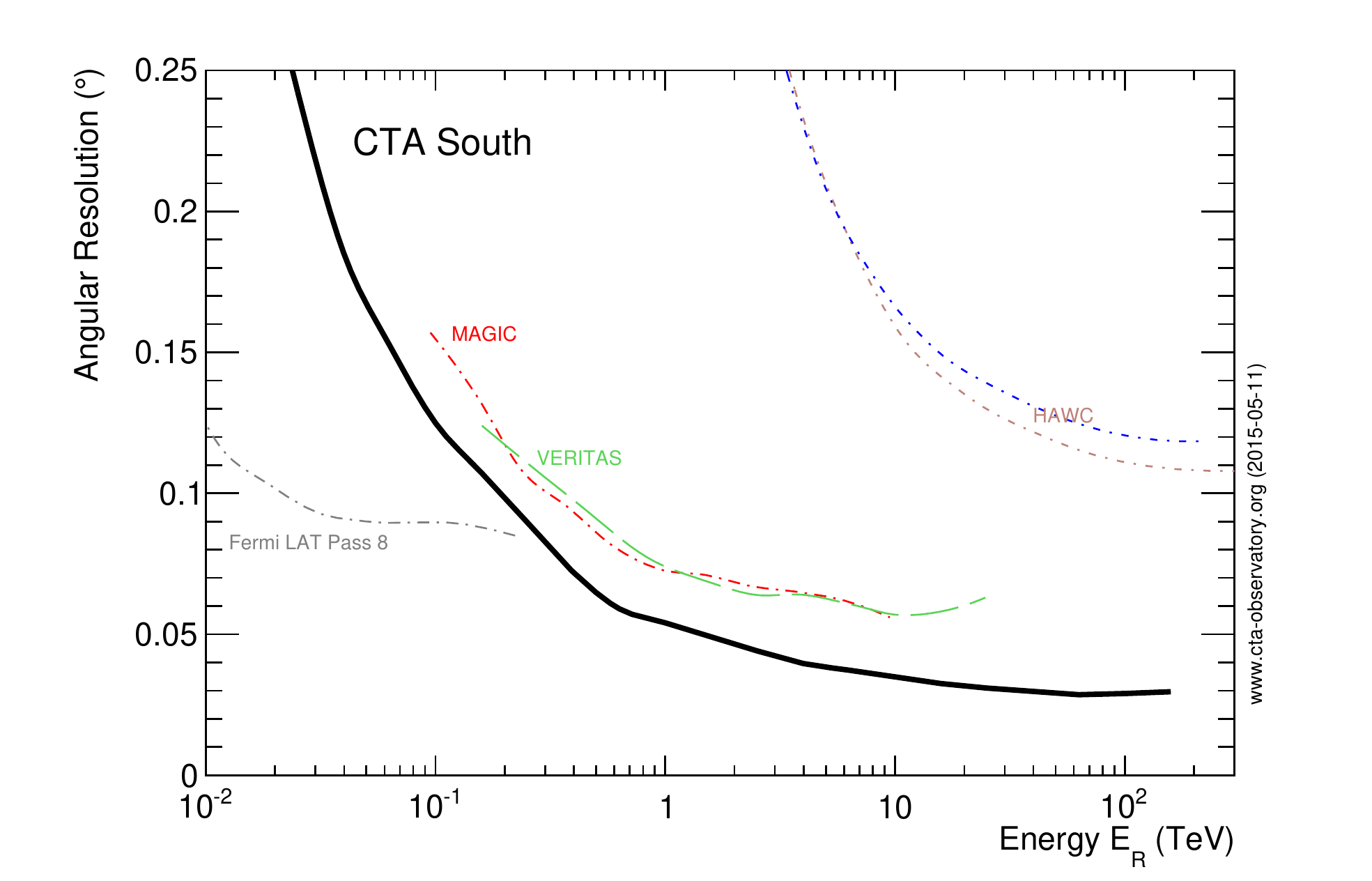}
    \caption{Angular resolution (68\% containment radius of the gamma-ray PSF) as a function of energy for CTA southern array (black line) \cite{CTAperf}. 
    For comparison angular resolutions for MAGIC-II (red line) \cite{MAGICperf}, 
    VERITAS (green line) \cite{VERITASperf}, HAWC (dark-red line) \cite{HAWCperf} and FERMI (gray line) \cite{FERMIperf} are shown.  
    }
  \label{fig:CtaAngRes}
\end{figure}

At the lowest end of the energy range, despite of the much larger collection area, the IACT sensitivity for an observation time of 50 hours is worse than that of satellite borne detectors, like Fermi, for one year observation time.
In this energy range such detectors achieve a background rejection much higher than IACT one, fully compensating for the smaller sensitive area. Their sensitivities are signal-limited in this energy range and scale linearly with the observation time, while IACT sensitivities are background-limited and therefore proportional to the square root of the observation time.      
The sensitivity of Cherenkov telescopes approaches that of satellite-borne detectors as the observation time decreases and 
eventually becomes even better. 
According to simulation, CTA will outperform the Fermi-LAT by 4-5 orders of magnitude in sensitivity for sub-minute-timescale transient phenomena such as certain AGN flares or GRBs.
The expected angular resolution of CTA is compared  
with that of some current and future gamma-ray detectors in figure \ref{fig:CtaAngRes}. 
The angular resolution of CTA will be better than that of any current IACT array over the full energy range but worse than the one of Fermi in the overlapping region. 
By selecting events observed by at least 10 telescopes the angular resolution could be further improved 
approaching an unprecedented 1 arc-minute at highest energies.   
  
\section{Scientific topics}
\label{CTAscience}

CTA project combines guaranteed scientific return, in the form of high precision astrophysics, 
with considerable potential for major discoveries in astrophysics and fundamental physics. 
The main questions that CTA will address can be roughly grouped in three main themes:
 
\subsection{Understanding the origin of cosmic rays and their role in the Universe}

One of the main goals of gamma-ray astrophysics has been to pinpoint sites of cosmic rays acceleration 
and to establish the dominant contributors to the locally measured flux of cosmic rays, which is to 99\% composed 
of protons and nuclei. 
Fermi and IACT arrays have identified the brightest accelerators in our galaxy providing 
strong evidence of hadron acceleration in some of them, but many key questions remain unanswered. 
It is not clear if supernova remnants are the only contributors to the galactic flux of cosmic rays, 
nor where particles can be accelerated to PeV energies in our galaxy.  
Detailed studies of galactic particle accelerators, such as pulsars and pulsar wind
nebulae, supernova remnants, and gamma-ray binaries are mandatory to answer such questions. 
CTA will address such questions exploiting its excellent surveying capability, extended energy coverage 
and better angular resolution.   


\subsection{Probing extreme environments}

Acceleration of particles up to very high energies is usually 
associated to extreme environments like the surroundings of compact
objects such as neutron stars and black holes. 
Micro-quasars at the galactic scale, and extragalactic sources like 
blazars, radio galaxies and Flat Spectrum Radio Quasars are the object of interest for this kind of investigations.  
CTA will be able to detect a large number of these objects thanks to its superior sensitivity 
and low energy threshold, enabling population studies which will be a major step forward in this area. 
Moreover, time-resolved spectral measurements and multi-wavelength campaigns will be the key to disentangle leptonic and
hadronic emission scenarios. 
Extragalactic background light (EBL), galaxy clusters and gamma-ray burst (GRB) studies are also connected to this topic.

\subsection{Exploring Frontiers in Physics}

CTA will conduct also searches for dark matter through
possible annihilation signatures, searches for axion-like particles, tests of Lorentz invariance violation at the Planck scale, 
and any other observational signatures that challenge our current understanding of fundamental physics.
The nature of dark matter is one of the most compelling questions presently facing physics and astronomy.
CTA indirect searches for dark matter will complement direct detection searches currently pursued by many other astro-particle experiments 
and searches carried on at particle accelerators.    

Photons with energies up to hundreds of TeV produced by distant cosmic sources represent a powerful tool 
to search for a wide range of new physics. 
Quantum gravity effects, for example, may induce time delays between photons with different
energies traveling over large distances due to a nontrivial refractive index of the vacuum. 
Observing a large number of extragalactic sources over a wide energy range CTA could disentangle 
this effect from a source intrinsic origin of the observed delay strongly constraining models about Lorentz Invariance Violation. 
 
Axions are expected to convert into photons and vice versa in the presence of magnetic fields. 
The Axion/photon coupling could modify the observed spectrum of distant sources. 
By detailed spectral studies of extragalactic sources CTA could calculate upper limits on Axion/photon coupling 
which will complement those from indirect astrophysical tests, from solar observations, and from X-ray telescopes.

%
%

\section{Summary}
\label{Summary}

The Cherenkov Telescope Array is a 
worldwide project for the next generation of ground-based 
Cherenkov Telescopes for Very High Energy Gamma ray astronomy.
The design of CTA telescopes, the layout optimization and the site selection
process are well advanced. The preparatory phase will be completed by the end of 2015 
and founding agreements are expected to be signed in 2016. 
The construction of first telescopes can start soon afterwards and the deployment of both arrays can be completed by 2020.
Early scientific results are expected already in 2017.
CTA will then be the major observatory
in VHE gamma-ray astronomy, combining guaranteed
astrophysics and physics returns with significant
discovery potential. Very significant synergies are expected with 
detectors operating in other energy ranges of the electromagnetic spectrum 
as well as with detectors exploiting other astrophysical messengers, 
like neutrinos and gravitational waves. 

\section{Aknowledgments}
\label{Akn}
We gratefully acknowledge support from the agencies and organizations
listed under Funding Agencies at this website: http://www.cta-observatory.org/.




\nocite{*}
\bibliographystyle{elsarticle-num}
\bibliography{martin}

\begin{thebibliography}{00}


\bibitem{CTA1} CTA website: https://portal.cta-observatory.org/Pages/Home\-.aspx

\bibitem{CTA2}  B. S. Acharia et al. (CTA Consortium), 
``Introducing the CTA concept'',
Astroparticle Physics {\bf 43} (2013) 3-18.  

\bibitem{HESS} H.E.S.S. website: https://www.mpi-hd.mpg.de/hfm/HESS/

\bibitem{MAGIC} MAGIC website: https://magic.mpp.mpg.de/

\bibitem{VERITAS} VERITAS website: http://veritas.sao.arizona.edu/


\bibitem{CTAscience} B. S. Acharia  et al. (CTA Consortium),  
``Seeing the High-Energy Universe with the Cherenkov Telescope Array. The Science Explored with the CTA'', 
Astroparticle Physics {\bf 43} (2013) 3-356. 

\bibitem{CTAproject} M. Actis  et al. (CTA Consortium), 
``Design Concepts for the Cherenkov Telescope Array'',   
Experimental Astronomy {\bf 32} (2012) 193-316. 

\bibitem{AGILE} AGILE website: http://agile.rm.iasf.cnr.it/

\bibitem{FERMI} FERMI website: http://fermi.gsfc.nasa.gov/

\bibitem{INTEGRAL} INTEGRAL website: http://sci.esa.int/integral/

\bibitem{WhippleCrab} T. C. Weekes et al, 
``Observation of TeV gamma rays from the Crab nebula using the atmospheric Cerenkov imaging technique'',
The Astrophysical Journal {\bf 342} (1989) 379-395.  

\bibitem{HESS_GalacticPlaneSurvey} C. Deil et al. (H.E.S.S. Collaboration), 
``The H.E.S.S. Galactic plane survey'', 
PoS(ICRC2015) 773 (2015)

\bibitem{HESS_RX_J1713} P. Eger et al. (H.E.S.S. Collaboration), 
``H.E.S.S. precision measurements of the SNR RX J1713.7-3946'', 
PoS(ICRC2015) 766 (2015)

\bibitem{S3_0218} J.Sitarek et al. (MAGIC Collaboration), 
``Detection of very-high-energy gamma rays from the most distant and gravitationally lensed blazar S3~0218+35 using the MAGIC telescope'', 
PoS(ICRC2015)825 (2015)

\bibitem{GammaRayReview} M. Lemoine-Goumard 
``Ground-based gamma-ray astronomy'', 
PoS(ICRC2015)012 (2015)   

\bibitem{DC} J. M. Davies and E. S. Cotton, 
``Design of the quartermaster solar furnace'', 
Solar Energy {\bf 1(2)} (1957) 16-22. 

\bibitem{SC} V. Vassiliev, S. Fegan and P. Brousseau,   
``Wide field aplanatic two-mirror telescopes for ground-based gamma-ray astronomy''
Astroparticle Physics {\bf 28} (2007) 10-27.   



\bibitem{HAWC} HAWC website: http://www.hawc-observatory.org/

\bibitem{LHAASO} LHAASO website: http://english.ihep.cas.cn/ic/ip/LHAASO/

\bibitem{TEVCAT} TevCat website: http://tevcat.uchicago.edu/

\bibitem{CTAsimulations} K. Bernl\"ohr et al. (CTA Consortium), 
``Monte Carlo design studies for the Cherenkov Telescope Array'', 
 Astroparticle Physics {\bf 43} (2015) 171-188. 
 
 
\bibitem{CORSIKA} D.Heck et al. Forschungszentrum Karlsruhe Report FZKA 6019 (1998)

\bibitem{SIMTELARRAY} K. Bernl\"ohr, 
``Simulation of Imaging Atmospheric Cherenkov Telescopes with CORSIKA and sim telarray'', 
Astroparticle Physics, {\bf 30} (2008) 149-158.  


\bibitem{CTAperf} CTA performance website: https://portal.cta-observatory.\-org/CTA\_Observatory/performance/SitePages/Home.aspx

\bibitem{MAGICperf} J. Aleksi\v{c} et al. (MAGIC Collaboration), 
``Performance of the MAGIC stereo system obtained with Crab Nebula data'', 
Astroparticle Physics {\bf 35} (2012) 435-448.

\bibitem{VERITASperf} VERITAS performance website: http://veritas.\-sao.\-arizona.\-edu/about-veritas-mainmenu-81/veritas-specifications-mainmenu-111

\bibitem{HAWCperf}
A. U. Abeysekara et al. (HAWC Collaboration)
``Sensitivity of the High Altitude Water Cherenkov Detector to Sources of Multi-TeV Gamma Rays''
Astroparticle Physics {\bf 50-52} (2013)  26-32. 

\bibitem{FERMIperf} Fermi-LAT performance website: http://www.slac.\-stanford.edu/\-exp/glast/groups/canda/\-lat\_Performance.htm

\end{thebibliography}



\end{document}